\documentclass[twocolumn,showpacs,preprintnumbers,amsmath,amssymb]{revtex4}
\usepackage[dvips]{graphicx}
\usepackage{hhline}

\begin{document}
\renewcommand{\abstractname}{Abstract}
\renewcommand{\refname}{References}
\renewcommand{\figurename}{Fig.}
\renewcommand{\tablename}{Table}
\newcommand{\bsig}{\mbox{\boldmath{$\sigma$}}}

\title{\center{Schiff moment of the Mercury nucleus and the proton dipole
moment}}
\author{V.F. Dmitriev and R.A. Sen'kov \\
{\it Budker Institute of Nuclear Physics, pr-t. Lavrentieva 11, Novosibirsk 90, 630090, Russia}\\
{\it and Novosibirsk State University, Pirogova st. 2, Novosibirsk
90, 630090, Russia}}

\begin{abstract}
We calculated the contribution of internal nucleon electric dipole
moments to the Schiff moment of  $^{199}$Hg. The contribution of
the proton electric dipole moment was obtained via core
polarization effects that were treated in the framework of random
phase approximation with effective residual forces. We derived  a
new upper bound $|d_p|< 5.4\times 10^{-24} e\cdot$cm of the proton
electric dipole moment.
\end{abstract}
 \maketitle
\section{Introduction}
 The interest to electric dipole
moments (EDM) of elementary particles and more complex systems like nuclei and
atoms  exists since 1950, when it was first suggested that there
was no experimental evidence for symmetry of nuclear forces under
parity transformation~\cite{pr50}. The interest was renewed after 1964 when it
was discovered that the invariance under CP transformation, which combines
charge conjugation with parity, is violated in $K$-meson decays. This provided a
new incentive for EDM searches. Since the combined CPT transformation is
expected to leave a system invariant, breakdown of CP invariance should be
accompanied by T violation. Thus, there is a reason to expect that P and T
violating EDMs should exist at some level.

The experimental upper limit on the neutron EDM is
\cite{pdg}
\begin{equation} \label{dn}
d_n < 0.63 \times 10^{-25}\;e\,{\rm cm}.
\end{equation}
The measured value for the proton EDM is   \cite{pdg}
\begin{equation}  \label{dp}
d_p = (-4 \pm 6) \times 10^{-23} \;e\,{\rm cm},
\end{equation}
and is compatible with zero.
This corresponds to an upper limit which is three order of magnitude weaker than
the one for the neutron.

The best upper limit on EDM ever obtained was in an atomic
experiment with $^{199}$Hg~\cite{rgf01}. The result for the dipole
moment of this atom is
\begin{equation} \label{exp}
d(^{199}{\rm Hg}) < 2.1 \times 10^{-28}\;e\,{\rm cm}.
\end{equation}
Unfortunately, the implications of this result  are
somewhat less impressive, due to the electrostatic screening of
the nuclear EDM in this essentially Coulomb system. The point is
that in a stationary state of such a system, the total electric
field acting on each particle must vanish. Thus, an internal
rearrangement of the system's constituents gives rise to an
internal field ${\bf E}_{int}$ that exactly cancels ${\bf E}_{ext}$ at each
charged particle; the external field is effectively switched off,
and an EDM feels nothing \cite{pr50,sch63, san67}.

Still, some P and T odd component of the electrostatic potential survives due to
final nuclear size. It is created by the next moment in the nuclear electric
dipole density distribution. This is the Schiff moment defined as  \cite{fks84}
\begin{equation} \label{1}
{\bf S} = \frac{1}{10} \sum_q e_q \left(r_q^2 {\bf r}_q -
\frac{5}{3}\langle r^2\rangle_{\mbox{ch}} {\bf r}_q\right).
\end{equation}
The Schiff moment generates a P and T odd electrostatic potential
in the form
\begin{equation} \label{2}
\phi({\bf r}) = 4\pi {\bf S}\cdot\mbox{\boldmath
$\nabla$}\delta({\bf r}).
\end{equation}
Interaction of atomic electrons with the potential given by Eq.
(\ref{2}) produces an atomic dipole moment
 \begin{equation} \label{3}
d_{atom} = \sum_n\frac{\langle 0| -e\sum_i^Z \phi({\bf
r}_i)|n\rangle\langle n|-e\sum_i^Z z_i|0\rangle}{E_n-E_0} + {\rm
H.c.}
\end{equation}
Because of the contact origin of the potential, only the electrons
in $s$ and $p$ atomic orbitals contribute to the dipole moment
given by Eq. (\ref{3}).

The Eq.(\ref{1}) is valid for any system of point-like charges $e_q$.
Let us split the sum in Eq.(\ref{1})  into the sum over coordinates of nucleons
and the sum over coordinates of charges inside the nucleons:
\begin{equation} \label{1.1}
{\bf S} = \frac{1}{10} \sum_N \sum_i  e_i \left(({\bf r}_N +
\mbox{\boldmath$ \rho$}_i )^2 - \frac{5}{3}\langle
r^2\rangle_{\mbox{ch}}\right) ({\bf r}_N + \mbox{\boldmath$
\rho$}_i).
\end{equation}
Here  ${\bf r}_N$ is a nucleon position and $ \mbox{\boldmath$
\rho$}_i$ is the position of the $i$th charge inside the nucleon.
Combining the terms of the zeroth and first order in $
\mbox{\boldmath$ \rho$} $ and using   $ \sum_i e_i = e_N$, $\sum_i
e_i \mbox{\boldmath$ \rho$}_i = {\bf d}_N$, we obtain an
expression for the Schiff moment as a sum of two terms. The first
of them is similar to (\ref{1})
\begin{equation} \label{1a}
{\bf S_1} = \frac{1}{10} \sum_N^A e_N \left(r_N^2 {\bf r}_N -
\frac{5}{3}\langle r^2\rangle_{ch} {\bf r}_N\right),
\end{equation}
where  $e_N$ is equal to $|e|$ for a proton and zero for a neutron.
The mean value of this operator is nonzero only in the
presence of the parity- and time-invariance violating nucleon-nucleon
interaction

The second term in the operator (\ref{1.1}) is related to the
internal dipole moments of the nucleons
\begin{eqnarray} \label{1b}
{\bf S_2} = \frac{1}{6} \sum_N^A {\bf d_N} \left(r_N^2 - \langle
r^2\rangle_{ch} \right) \nonumber \\
+\frac{1}{5} \sum_N^A \left( {\bf r}_N ({\bf r}_N \cdot
{\bf d}_N) - {\bf d}_N r_N^2/3 \right).
\end{eqnarray}

The previous calculations of the Schiff moment of a heavy
nucleus \cite{fks84,fg02} were performed in a simplified manner,
without taking into account the residual interaction between a valence nucleon
and the core nucleons. Only recently more microscopic studies of the Schiff
moment of $^{199}$Hg \cite{ds03} and $^{225}$Ra \cite{ebd03} appeared where
effects of the core polarization with the effective forces for
$^{199}$Hg and the octupole deformation for $^{225}$Ra based on
Skyrm-Hartree-Fock method were discussed. In this work we would
like to concentrate on the nucleon EDM contribution to the Schiff
moment of the $^{199}$Hg nucleus. In the picture of independent particle
model only an EDM of a valence nucleon contributes to the Schiff
moment. In case of $^{199}$Hg it is a neutron EDM. However, when a
residual quasiparticle interaction between the valence neutron and the protons
in the core is taken into consideration, the proton EDM
contribution to the nuclear Schiff moment becomes nonzero. We calculated this
contribution using a random phase appriximation with effective forces. From
the relation between the Schiff moment and the electric dipole moment of the Hg
atom \cite{dfgk}  the new upper limit on the proton EDM was obtained.

\section{Outline of the theory}

\subsection{Nuclear mean field}
In our calculations we used full single-particle spectrum
including continuum. The single-particle basis was obtained using
partially self-consistent mean-field potential of  \cite{bs74}.
The potential includes four terms. The isoscalar term is the
standard Woods-Saxon potential
\begin{equation} \label{6}
U_0(r) = - \frac{V}{1+\exp{\frac{r-R}{a}}},
\end{equation}
with the parameters $V = 52.03$ MeV, $R = 1.2709 A^{1/3}$ fm, and
$a = 0.742$ fm. Two other terms  $U_{ls}(r)$, and $U_\tau(r)$ were
obtained in a self-consistent way using two-body Landau-Migdal-type
interaction of   \cite{m67} for the spin-orbit and isovector parts
of the potential. The last term is the Coulomb potential
 of a uniformly charged sphere  with $R_C = 1.18 A^{1/3}$
fm. The mean field potential obtained in this way produces a good
fit for single particle energies and r.m.s. radii for nuclei in
the region around $^{208}$Pb.

\subsection{Core polarization}
The effects of the core polarization for a single particle
operator can be treated by introducing a renormalized operator
$\tilde{{\bf S}}$ satisfying the equation
\begin{equation} \label{11}
\tilde{\bf S}_{\nu'\nu}  = {\bf S}_{\nu'\nu} + \sum_{\mu'\mu}
\tilde{\bf S}_{\mu\mu'} \frac{n_\mu -n_{\mu'}}{\epsilon_\mu -
\epsilon_{\mu'}+\omega} \langle \nu'\mu'|F|\mu\nu\rangle,
\end{equation}
where ${\bf S}$ is the bare Schiff moment operator given by
(\ref{1}), (\ref{1a}) or (\ref{1b}). $n_\mu$  and $\epsilon_\mu$
are  single particle occupation numbers and energies.  For
static moments the external frequency $\omega \rightarrow 0$.
The value of the Schiff moment is given by the diagonal matrix element of the
z-component of the renormalized operator (\ref{11}) between mean field states of
the last unpaired nucleon with a maximal angular momentum projection.
\begin{equation} \label{12}
 S = \langle \mu j\,m=j|\tilde{\bf S}_z|\mu j\,m=j \rangle.
\end{equation}

For the residual interaction $F$ we use the phenomenological Landau-Migdal
interaction that has the form
\begin{equation}\label{15}
F = C\left[g_s(\mbox{\boldmath $\sigma$}_1\cdot \mbox{\boldmath
$\sigma$}_2) + g'_s (\mbox{\boldmath $\sigma$}_1\cdot
\mbox{\boldmath $\sigma$}_2) (\mbox{\boldmath $\tau$}_1\cdot
\mbox{\boldmath $\tau$}_2)\right]\delta({\bf r}_1-{\bf r}_2),
\end{equation}
where $C=$ 300 MeV fm$^3$.  The values of the empirical
interaction constants $g_s$ and $g'_s$ are crucial for our
calculations. The proton contribution is proportional to the
proton - neutron interaction $g_s - g'_s$. The constant $g'_s$ is
determined from magnetic properties of nuclei and positions of
Gamow-Teller resonances. Its adopted value varies between
$g'_s=0.9$--$1$ depending on details of the mean field potential
used \cite{sww77,dt83, FrOst}. The constant $g_s$ is not so well
defined. The magnetic moments and M1 transitions are to large
extent isovector and they do not fix $g_s$.  An attempt to fix it
from the structure of high spin states in $^{208}$Pb has been done
in \cite{ks80}. They found that $g_s=0.25$ had to be used in order
to reproduce the excitation energies of the $M12$ and  $M14$
states. Another value $g_s=0.19$ was quoted in the review paper
\cite{FrOst}.

The Schiff moment operator can be presented in coordinate space
in the form
\begin{equation}
S_{1 m} = \,\sum_{i=1}^{2}\,S^{\;i}(r)\,T^{(i)}_{1 m},
\end{equation}
where we have introduced the  set of linear independent
tensor operators
\begin{equation}
\begin{array}{c}\label{operators}
T^{(1)}_{JM}\,=\,\bsig \cdot {\bf Y}^{J-1}_{JM}({\bf n}),\;
T^{(2)}_{JM}\,=\,\bsig \cdot {\bf Y}^{J+1}_{JM}({\bf n}),\\
\end{array}
\end{equation}
where ${\bf Y}^L_{JM}({\bf n})$ is the vector spherical harmonic. For $J=1$ we
have $T^{(1)}_{1m}\sim  \sigma_m $, and $T^{(2)}_{1m}\sim n_m ({\bf n}\cdot
\mbox{\boldmath $\sigma$})-\frac{1}{3}\sigma_m$. For a
spherical nucleus we can separate the angular variables and solve the obtained
equations in coordinate space. The equations are
\begin{equation}\label{16}
S^{a i}(r) = S_0^{a i}(r) +\int_0^\infty A^{ai \; bj}(r,r')
S^{b j}(r')\,dr' ,
\end{equation}
where  $a=p,n$ and  $S_0^{a i}(r)$  is the radial part of the
Schiff moment operator (\ref{1b}) multiplied by $r$. The kernel of the integral
equation $A^{a i\; b j}(r,r')$ was calculated by means of the
Green functions of the radial Schr\"odinger equation.
$$
A^{a i\;b j}(r,r') = \frac{C g^{ab}}{3}
\sum_{\kappa jl} n^b_\kappa \langle jl||T^{(i)}_1||\kappa\rangle \;
\langle jl||T^{(j)}_1||\kappa\rangle rR^b_\kappa(r)
$$
\begin{equation}\label{17}
\times r'R^b_\kappa(r') \left( G^b_{jl}(r,r'|\epsilon_\kappa+\omega) +
G^b_{jl}(r,r'|\epsilon_\kappa -\omega)\right),
\end{equation}
where $g^{pp} = g^{nn}=g_s + g'_s,\; g^{pn}=g^{np}=g_s-g'_s$, $R^b_\kappa (r)$
are the radial wave functions, and $n^b_\kappa$ are the occupation numbers.

The solutions of Eq. (\ref{11}) for $S^{ai}(r)$ are shown in Fig.1
and 2.
\begin{figure}
\includegraphics[width=8cm]{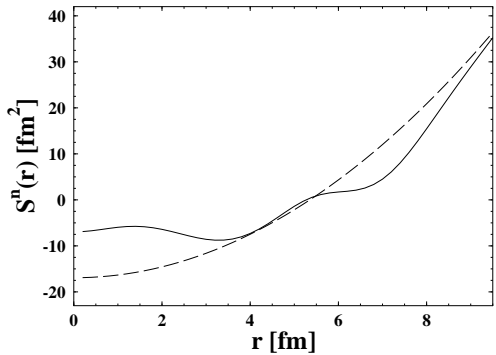}
\caption{Core polarization effects in the neutron Schiff moment
operator. The solid curve is $S^{n1}(r)$; the dashed curve is the
bare operator $S_0^{n1}(r)$.}
\end{figure}
\begin{figure}
\includegraphics[width=8cm]{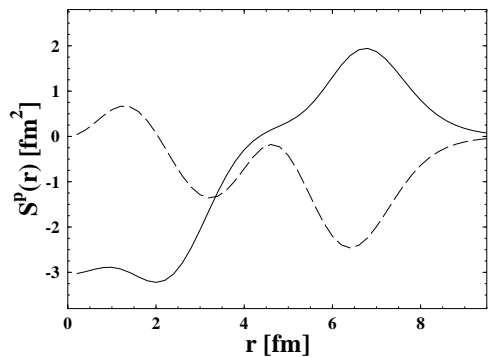}
\caption{Radial dependence of the proton effective Schiff moment
operators induced by the core polarization. The solid curve is
$S^{p1}(r)$; the dashed curve is $S^{p2}(r)$.}
\end{figure}
Figure 1 demonstrates the magnitude of the core polarization
effects. Repulsive residual interaction (\ref{15}) leads to not
very significant decrease of the mean value of the Schiff moment.
Figure 2 shows the radial dependence of the proton contributions
induced by the core polarization. Note 1 order of magnitude
difference in the scales in Fig. 1 and 2. The full curve in Fig. 2
is the radial dependence at the first operator $T^{(1)}_{1m}$ and
the dashed curve is the radial dependence at the second operator
$T^{(2)}_{1m}$. $S^{p1}(r)$ changes sign inside the nucleus,
therefore its mean value is smaller than the mean value of
$S^{p2}(r)$ which is mostly negative inside the nucleus.

\section{Results}
The value of the Schiff moment of $^{199}$Hg can be presented as a sum of proton
and neutron contributions
$$
S = s_pd_p + s_n d_n.
$$
In Table I we list the values $s_p$ and $s_n$ calculated for
different combinations of $g_s$ and $g'_s$.
\begin{table}
\caption{Values of $s_p$ and $s_n$ for different $g_s$ and
$g'_s$.}
\begin{ruledtabular}
\begin{tabular}{c|c|c|c}\label{tabr}
$s_p$ & $s_n$ & $g_s$ & $g'_s$ \\
\hline
0.18 &  1.89 & 0.25 &  0.9 \\
0.19  &  1.86 & 0.25 & 1.0\\
0.20  &  1.93 & 0.19 & 0.9 \\
0.22  &  1.90  & 0.19 & 1.0\\
\end{tabular}
\end{ruledtabular}
\end{table}
From Table \ref{tabr} one can see that the uncertainties in $s_p$
and $s_n$ due to uncertainties in $g_s$ and $g'_s$ are
\begin{equation} \label{spsn}
s_p = 0.20 \pm 0.02\,\mbox{fm}{}^2,\;\; s_n= 1.895\pm 0.035\,\mbox{fm}{}^2.
\end{equation}
The main contribution to $s_p$ and $s_n$ comes from the second term in Eq.
(\ref{1b}).  The contribution of the first term is only -0.7 fm$^2$ in $s_n$ and
0.006 fm$^2$ in $s_p$.

The constraint for the Schiff moment of Mercury nucleus  from the experiment \cite{rgf01}
can be obtained using the results of Ref.
\cite{dfgk}. They calculated EDM of an atom created by the nuclear Schiff
moment. For $^{199}$Hg they found
$$
d = -2.8 \times 10^{-17} S\;[\rm e\,fm^3].
$$
From Eq.(\ref{exp}) we obtain the following upper bound for the Schiff moment
 \begin{equation}\label{sm}
 |S({}^{199} \mbox{Hg})| < 0.75 \times 10^{-11} \;
\mbox{e fm}{}^3,
\end{equation}
From Eq. (\ref{sm}) we can give the following constraints for EDM of nucleons:
\begin{equation}\label{an1}
|d_p| < 3.8 \times 10^{-24} \; \mbox{e cm,} \;\; |d_n| < 4.0
\times 10^{-25} \; \mbox{e cm}.
\end{equation}
The constraint for the neutron EDM is worse than the existing
result  $ d_n < 0.63 \times 10^{-25} \; \mbox{e cm}$ \cite{pdg},
therefore, we shall not discuss it below.  For proton EDM the
estimate (\ref{an1}) is 1 order of magnitude lower than the
existing experiment $ d_p = (-4 \pm 6) \times 10^{-23}$ e cm
\cite{pdg}. In these circumstances the question about a real
theoretical accuracy of our approach becomes important. It is
clear that the value $\pm 0.02$ cited in Eq.(\ref{spsn}) does not
reflects the real accuracy of the theory. It just came from the
difference in adopted values of $g_s$ and $g'_s$. The theoretical
uncertainty appears from two sources. First, it is an uncertainty
in the atomic calculations that couple the nuclear Schiff moment
and EDM of an atom. We shall not discuss it here referring to the
work \cite{dfgk}. Second, it is an uncertainty in calculations of
the core polarization effects using RPA with effective forces. The
latter can be estimated from the following considerations. Using
RPA with the effective forces we can fit different nuclear moments
in one nucleus. Then, in neighbor nuclei the calculated moments
will differ from the data. This difference can be regarded as an
uncertainty in the theory. In our experience this difference is of
the order of 20\% on the average, reaching sometimes the value of
30\% \cite{ds02}. To be safe, we can adopt a conservative 30\%
uncertainty in calculations of $s_p$. Therefore, instead of
(\ref{spsn}) we would prefer to write for $s_p$
\begin{equation} \label{sp}
s_p = 0.2 \pm 0.06 \, {\rm fm}^2.
\end{equation}
Since the error in (\ref{sp}) is not statistical, we can not give
a probability distribution for $s_p$. If one takes $0.14\,{\rm fm}^2$ as a
minimal value of $s_p$, then
 it gives the following  value for the proton
EDM upper bound
\begin{equation}
|d_p| <  5.4\times 10^{-24} \, e\cdot {\rm cm}.
\end{equation}

In summary, we calculated the contributions of the proton and
neutron EDM to the Schiff moment of  $^{199}$Hg. The effects of core
polarization were accounted for in the scope of RPA with the
effective residual forces. A new upper bound of the proton EDM has
been obtained from the upper bound on the atomic EDM of $^{199}$Hg
atom.

\begin{acknowledgements}
The authors are grateful to I. B. Khriplovich for bringing our
attention to this problem and for useful discussions.
\end{acknowledgements}


\begin{thebibliography}{99}
\bibitem{pr50} E.M. Purcell and N.F. Ramsey, Phys. Rev.  {\bf 78}, 807
(1950).
\bibitem{pdg} K. Hagiwara {\it et al.}, Phys. Rev. D {\bf 66}, 010001 (2002).
\bibitem{rgf01} M.V. Romalis, W.C. Griffith, and E.N. Fortson, Phys. Rev. Lett.
 {\bf 86}, 2505 (2001).
\bibitem{sch63} L.I. Schiff, Phys. Rev.  {\bf 132}, 2194 (1963).
\bibitem{san67} P.G.H. Sandars, Phys. Rev. Lett.  {\bf 19}, 1396 (1967).
\bibitem{fks84} V.V. Flambaum, I.B. Khriplovich, and O.P. Sushkov Zh. Eksp.
Teor. Fiz.  {\bf 87}, 1521 (1984) [Sov. Phys. JETP  {\bf 60}, 873 (1984)].
\bibitem{fg02} V.V. Flambaum and J.S.M. Ginges, Phys. Rev. A {\bf 65},
032113 (2002).
\bibitem{ds03} V.F. Dmitriev and R.A. Sen'kov, arXiv:nucl-th/0304048.
\bibitem{ebd03} J. Engel, M. Bender,  J. Dobaczewski, J.H. de Jesus, and P.
Olbratowski, arXiv:nucl-th/0304075.
\bibitem{dfgk} V.A. Dzuba, V.V. Flambaum, J.S.M. Ginges, M.G. Ko
zlov, Phys. Rev. A {\bf 66} 012111 (2002).
\bibitem{bs74} B.L. Birbrair and V.A. Sadovnikova, Yad. Fiz.
{\bf 188}, 1851 (1974).
\bibitem{m67} A.B. Migdal, {\it Theory of Finite Fermi System},
(Wiley, New York, 1967).
\bibitem{sww77} J. Speth, E. Werner and W. Wild, Phys. Rep. {\bf 33}, 127
(1977).
\bibitem{dt83} V.F. Dmitriev and V.B. Telitsin, Nucl. Phys. {\bf A402}, 581
(1983) \bibitem{FrOst} F. Osterfeld, Rev. Mod. Phys. {\bf 64}, 491 (1992).
\bibitem{ks80} S. Krewald and J. Speth, Phys. Rev. Lett. {\bf 45}, 417 (1980).
\bibitem{ds02} V.F. Dmitriev, R.A. Sen'kov, Nucl. Phys. A {\bf 706},
351 (2002).
\end{thebibliography}
\end{document}